\documentclass[preprint2,a4paper]{aastex}

\usepackage{epstopdf}
\usepackage{emulateapj5}
\usepackage{onecolfloat5}

\newcommand{\Spk}{S_{\rm pk}}
\newcommand{\be}{\begin{eqnarray}}
\newcommand{\ee}{\end{eqnarray}}

\newcommand{\vs}{vs.\ }

\input{crab.defs}

\shortauthors{Bhat et al.}
\shorttitle{Arecibo Search}
\begin{document}
\twocolumn[
\title{An Arecibo Search for Pulsars and Transient Sources in M33}
\medskip
\author{N. D. R. Bhat$^1$, J. M. Cordes$^2$, P. J. Cox$^1$, J. S. Deneva$^2$, T. H. Hankins$^3$, T. J. W. Lazio$^{4,5}$, M. A. McLaughlin$^{6,7}$ }
\affil{$^1$Centre for Astrophysics \& Supercomputing, Swinburne University, Hawthorn, Victoria 3122, Australia \\
$^2$Astronomy Department, Cornell University, Space Sciences Bldg, Ithaca, NY 14853, USA \\
$^3$Physics Department, New Mexico Tech, Socorro, NM 87801 \\
$^4$Jet Propulsion Laboratory, California Institute of Technology, Pasadena, CA  91109  USA\\
$^5$SKA Program Development Office, Alan Turing Building, Oxford Road, Manchester, M13 9PL, UK \\
$^6$Department of Physics, West Virginia University, Morgantown, WV 26506, USA \\
$^7$Also adjunct at National Radio Astronomy Observatory, Green Bank, WV 24944, USA}
\begin{abstract}
We report on a systematic and sensitive search for pulsars and transient sources in the nearby spiral galaxy M33, 
conducted at 1.4~GHz  with the Arecibo telescope's seven-beam receiver system, ALFA. 
Data were searched for both periodic and aperiodic sources, up to 1000 \dmu\,in dispersion measure
and on timescales  from $\sim$50 \us to several seconds. 
The galaxy was sampled with 12 ALFA pointings, or 84 pixels in total, each of which was searched for 2--3 hr. 
We describe the observations, search methodologies and analysis strategies applicable to multibeam systems, and 
comment on the data quality and statistics of spurious events that arise due to radio frequency interference. 
While these searches have not led to any conclusive signals of periodic or transient nature that originate in the galaxy, 
they illustrate some of  the underlying challenges and difficulties in such searches and the efficacy of simultaneous 
multiple beams in the analysis of search output. 
The implied limits are $\la$5 \lumuu\,in luminosity (at 1400 MHz) for periodic sources in M33 with duty cycles $\la$5\%.  
For short-duration transient signals (with pulse widths $\la$100 \us), the limiting peak flux density is 100~mJy, 
which would correspond to a 5-$\sigma$ detection of bright giant pulses ($\sim$20 kJy) from Crab-like pulsars if 
located at the distance of M33.
We discuss the implications of our null results for various source populations within the galaxy and comment on the 
future prospects to conduct even more sensitive searches with the upcoming next-generation instruments including
the Square Kilometer Array and its pathfinders. 
\end{abstract}
\keywords{galaxies: individual (M33) -- pulsars: general -- pulsars: individual (Crab pulsar, PSR B0045+33, PSR J0628+0909) --  surveys}
]

\section{Introduction} \label{s:intro}

Numerous pulsar surveys over the past four decades have led to an impressive tally of $\sim$2000 pulsars known within our Galaxy  \citep{manchesteretal2005}. The Magellanic clouds, located at a distance of $\sim$50 kpc, are known to harbour 20 pulsars \citep{manchesteretal2006}, which are the most distant pulsars currently known. Finding pulsars in galaxies other than our own still remains a major challenge, as the achievable sensitivities of most currently operational radio telescopes are well below that required to detect even the most luminous pulsars in nearby galaxies. Nonetheless, discoveries of such extragalactic pulsars will be scientifically rewarding on multiple counts: in addition to providing valuable insights into the pulsar populations and birth rates within the host galaxies, such pulsars, through their dispersion measures (DMs) and rotation measures (RMs), have the potential to probe the baryon density and magnetic field in the intervening intergalactic medium as well as the interstellar medium of the galaxies in which they reside \citep{clm04,cordes2007}. 

Of all the existing instruments, the Arecibo 305-m radio telescope is the most promising to conduct such searches given its superior sensitivity. However, its inherent limitation in sky coverage restricts the number of suitable targets. Within the Local Group of galaxies, 
the Triangulum galaxy M33, a spiral galaxy located at 0.95 Mpc \citep{vivianetal2009,bonanosetal2006}, is a highly promising target. Its location away from the Galactic plane ($ | b | \sim 30^{\circ} $) and the face-on orientation mean only moderate values of dispersion measure are expected due to Galactic or extragalactic material and therefore the search is unlikely to be limited by dispersion or scattering
at frequencies above $\sim$1 GHz.  Previous Arecibo searches by \citet{mc03} were made at a frequency of 430 MHz and targeted giant-pulse like signals over a large DM range of 0--250 \dmu.

As highlighted by \citet{cm03}, standard periodicity searches hold limited prospects for finding pulsars out to such large distances. However, alternate strategies, such as the detection of giant pulses from prospective Crab-like pulsars in the galaxy, seem
promising \citep{cm03,mc03,cbh04}. Giant pulses are sporadic bursts of emission with amplitudes reaching as large as two to four orders of magnitude brighter than those of regular pulses. For the Crab pulsar, they are known to reach amplitudes of $\sim$160 kJy at 430 MHz \citep{cbh04} and $\sim$45 kJy at 1.4 GHz \citep{bhatetal2008}.  Such ``giant-pulse'' emission, once thought to be unique to the Crab, has now been detected from several more objects, including millisecond pulsars (MSPs) B1937+21 and J1823$-$3021A \citep{cognardetal1996,knightetal2005} and the LMC pulsar B0540$-$69 \citep{jr2003}. 
The estimated brightness temperatures are known to reach up to $10^{41}$\,K for sub-ns giant pulses from the Crab 
\citep{hankins2007} and $10^{39}$\,K for the MSP B1937+21 \citep{soglasnovetal2004}.

One of the distinguishing characteristics of giant pulses is that their pulse amplitudes follow a  power-law distribution \citep{lundgrenetal1995}. 
However, no upper amplitude cutoffs have ever been observed for any of the objects studied so far \citep{lundgrenetal1995,soglasnovetal2004,knight2006}, implying that exceedingly bright pulses are potentially detectable if observations are made over sufficiently long durations. 
Unlike periodicity searches, search strategies of this kind do not
require unrealistically long integration times per pointing. However,
they rely heavily on the chance occurrence of such ultra-bright pulses
during observations and consequently their expected rates of occurrence.

\citet{cm03} discussed the methodologies and strategies relevant for searching for individual, dispersed pulses in fast-sampled spectrometer data. While originally developed for searching for giant pulse-like signals in the data, these techniques are fairly generic and hence applicable for searches of other targets that may emit transient signals. \citet{mc03} report on systematic searches conducted with the Arecibo and Parkes telescopes toward several targets including M33 and the LMC. Subsequent applications of these transient search techniques led to the discoveries of rotating radio transients \citep[RRATs;][]{mclaughlinetal2006} and a powerful  burst of unknown origin \citep{lorimeretal2007} and, most recently, several more RRAT-like objects in the PALFA survey data \citep{denevaetal2009} and in the 
Parkes survey data \citep{keane2010}. 

While the searches for extragalactic pulsars by  McLaughlin \& Cordes (2003) did not result in any convincing detections, they highlighted some of the underlying difficulties and challenges involved in such searches. 
In particular radio frequency interference (RFI) poses a major impediment in such searches. 
Impulsive and powerful RFI bursts can potentially subdue, or even mask, the dispersed radio pulses from pulsars, and often tend to manifest as spurious signals at non-zero DMs. \citet{mc03} found several tantalizing candidates in their data, however the single-pixel nature of their observations meant limited capability to distinguish them from spurious signals of RFI origin, and therefore greatly limited their ability to conclusively establish a plausible astrophysical origin for any of them.  
In a follow-up work, \citet{bhatetal2005} conducted exploratory searches toward M33 using simultaneous observations made with Arecibo and and the Green Bank Telescope (GBT). While the total observing time was rather too small and the sensitivity of the GBT too low to provide any detections, they demonstrated the efficacy of the {\em dual-site} aspect as a means of identifying and classifying various kinds of RFI (including that local to one or the other telescope and also satellite interference that is common to both). 
In any case, persistent RFI signals may adversely impact the detection sensitivities, as they tend to raise apparent detection thresholds, which means reduced sensitivities to weaker signals. 


\cite{lorimeretal2007} reported a strong pulse near the direction of the Small Magellanic Cloud (SMC)
which had a dispersion measure (DM) too high to originate from the Milky Way or from the SMC.    
More recently, \citet{bsetal2010} argued that it was swept-frequency RFI  that convincingly 
approximated the expected characteristics of an extragalactic radio transient.
Regardless of such details and the nature of its origin, the report has stimulated 
great interest in probing the transient radio sky on sub-second time scales. 


In \S~2 we discuss the feasibility of using the Arecibo telescope to
search for periodic and transient signals from M33.  \S~3 describes
the data acquisition, \S~4 the processing for transients and \S~5 the
periodicity search.   We analyze the results in terms of source
populations in \S~6 and summarize future possibilities in \S~7.  Our
conclusions are presented in \S~8.

\section{Arecibo Search Sensitivity and Source Detectability} \label{s:search}

Arecibo's seven-pixel receiver system ALFA (Arecibo L-band Feed Array) offers an exciting opportunity to revisit such searches. Specifically, the combination of high sensitivity, large bandwidth and multiple beams provides
both large instantaneous sky coverage as well as a powerful discrimination against spurious signals 
of RFI origin. Since  2004, the system has been exploited for multiple large
surveys including those for pulsars and short-duration transients \citep{cordesetal2006,denevaetal2009}. In this paper, we report on the first systematic searches conducted with the ALFA system toward the nearby spiral galaxy M33. Our searches targeted both periodic and aperiodic sources over a wide range of parameter space. Data quality is found to be remarkably good, with only a small fraction of the data corrupted by RFI. We discuss the efficacies of search strategies and methodologies that take advantage of ALFA's multiple beams and discuss the implications of our null results for various classes of source populations in the galaxy and place upper limits on their luminosities. 

Before we delve into the details of observations and data processing we briefly discuss different types of sources and signals that our searches are potentially sensitive to and comment on their detection prospects  and the achievable sensitivities at a distance of $\sim$1 Mpc. These include: 
(i) ultra-luminous pulsars, 
(ii) bright giant pulses from Crab-like pulsars, 
(iii) rotating radio transients that emit ultra-bright bursts,
(iv) transient radio emission from magnetar-like objects, detectable either as individual pulses or through periodic emission, and
(v) other exotic bursts of Galactic or extragalactic origin.  
While our survey was originally motivated by (i) and (ii), our processing techniques and strategies
are fairly general and thus target a much broader range of objects including the source classes of
(iii) to (v). 

\begin{figure}[t]
\epsscale{1.00}
\plotone{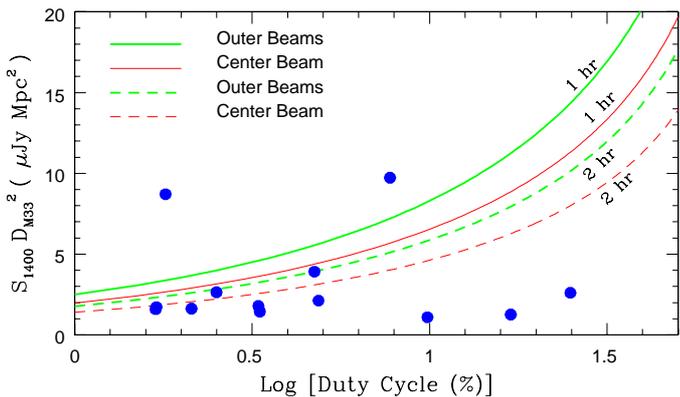}
\caption{Sensitivity curves for periodicity searches with the ALFA system, 
plotted as minimum detectable pulsar luminosities at a distance of 
$\sim$ 1 Mpc for a 8-$\sigma$ detection threshold; the solid and 
dashed lines correspond to 1 hr and 2 hr integrations, respectively; with the 
thick (red) and thin (green) lines corresponding to the center and 
outer beams. Observing bandwidth assumed is 100 MHz and the center 
frequency is 1.4 GHz. The filled circles denote the 12 highest 
luminosity pulsars known within our Galaxy ($ L _{1400} \protect\ga 10^3 ~ \lumu $), 
if they were located at the distance of M33 
($ D_{\rm M33} \sim $ 1 Mpc).} \label{sens}
\end{figure}

\subsection{Periodicity Search} \label{s:periodicity}

For periodicity searches, the achievable sensitivity  with ALFA is given by a modified form of the radiometer equation:
\begin{eqnarray}
{ S \over N} &  =  & \left( { \Lav \over 0.5\, \lumuu } \right) \left( { \Dgal \over D }  \right)^2 \left( { \Ssys \over 3\,{\rm Jy} } \right)^{-1} \nonumber \\
& &  \left[ \left( { \delnu \over 100\,{\rm MHz} } \right) \left( { \Tint \over 2\,{\rm hr} } \right) \left( { \Npol \over 2 } \right) \right]^{1/2} \left( { { \Wp / P } \over 0.05 } \right)^{-1/2} 
\label{eq:per}
\end{eqnarray}
\noindent
where \Lav is the period-averaged  pseudo-luminosity at 1.4 GHz (L band) in units of \lumuu; given by $ \Sav D^2$, where \Sav is the period-averaged flux density and $D$ the distance; \Dgal is the distance to M33. 
We note that 1~\lumuu is equal to $10^3$~mJy~kpc$^2$, near the high end of pseudo-luminosities of Galactic pulsars at 1.4~GHz. 
Other quantities are:
P, the rotation period; \Wp, the measured pulse width (thus \Wp/P is the duty cycle); S/N, the signal-to-noise ratio;  \Ssys, the system sensitivity in Jy (given by \Tsys/G where  \Tsys is the system temperature and G is the telescope gain in \kpjy); \delnu and \Tint, the total bandwidth and integration time over which the data are recorded; and \Npol, the number of  polarizations.
The distance to M33 was previously estimated to be approximately 840 kpc \citep{freedmanetal1991,freedmanetal2001}, but has recently been revised upward to 950 kpc \citep{vivianetal2009,bonanosetal2006}. For the nominal sensitivity parameters of the ALFA system (see \citet{cordesetal2006}), 
\Tsys= 30 K and G=10.4 \kpjy on-axis (i.e., center beam) and 8.2 \kpjy off-axis (i.e., outer beams); hence, \Ssys = 3.7 Jy and 3 Jy as the theoretical values for the outer and center beams respectively. 

The achievable sensitivities are shown in Figure~\ref{sens} where the minimum detectable luminosity is plotted against the
duty cycle of the pulsed emission, along with the $\sim$10 most luminous pulsars known within our Galaxy ($\Lav \ga 1 \, \lumuu$). 
Of the 1880 pulsars known so far, nearly 75\% have \Lav measured at 1400 MHz. Interestingly, two pulsars -- PSR J1644$-$4559 and PSR J0738$-$4042 with \Lav in the 6--8 \lumuu range -- will be well above the predicted detection curves (assuming 1 hr integration
and 100 MHz recording bandwidth) even when moved to a distance of 1 Mpc, with marginal detections possible for two others for longer (2 hr) integration.  For an 8-$\sigma$ detection threshold, this corresponds to a minimum period-averaged flux density \Sav $\sim$ 4 \uJy in a 2 hr integration. So, if M33 harbors a pulsar population with a similar luminosity distribution, indeed there should be at least a few pulsars that are potentially detectable with ALFA's sensitivity and integration times of $\sim$2 hr.

\bigskip

\bigskip

\subsection{Transient Search} \label{s:transient}

The search sensitivity for transient objects is given by a simpler form of the above radiometer equation,
\begin{eqnarray}
{ S \over N} &  = & \left( { \Dgal \over D } \right)^2 
\left( { \Spk \over { 20 \rm \, mJy } } \right) 
\left( { \Ssys \over 3\,\rm Jy } \right)^{-1} \nonumber \\
 & & \left[ \left( { \delnu \over 100\,\rm MHz } \right) \left( { \Wp \over 0.1\, \rm ms  } \right) \left( { \Npol \over 2 } \right) \right]^{1/2} 
 \label{eq:trans}
\end{eqnarray}
where $\Spk$ is the peak flux density and \Wp is the measured width of the detected pulse. 
This measured width is determined by the intrinsic pulse width, scatter broadening from 
multipath propagation, and dispersive smearing due to the instrument and any error in DM. 
Assuming \Wint, \Wscatt, \Wdm and \Wdmerr denote these terms, the net pulse width measured is given by 
\begin{equation}
W^2  = \Wintsq + \Wdmsq + \Wdmerrsq + \Wscattsq 
\end{equation}
Toward M33's direction ($ l = 133.6^{\circ}$, $ b = -31.3^{\circ}$), the predicted Galactic contribution to the dispersion measure is only $\sim$50 \dmu \citep{ne2001} and therefore the pulse broadening is negligible at frequencies of 1--2 GHz (\Wscatt $ \sim 0.1 \, \us $). Even if the model prediction is off by a factor of two,  the  dispersion smearing (\Wdm) expected is $ \sim $100 $ \mu $s for a frequency channel width of 400 kHz. Thus, pulses that are intrinsically very narrow (\Wint $\la$ 100 \us) will be seen as $\sim$100 \us pulses, while broader ones will encounter little smearing from dispersion or scattering along the line of sight.
 
Based on the above equation, a giant pulse that is as bright as 20 kJy at the Crab's distance ($\sim$2 kpc) will be potentially detectable at M33's distance ($\sim$1 Mpc) at the 5-$\sigma$ level (i.e., a peak flux density, $\Spk \sim$100 mJy). Thus, detection prospects critically depend on the rates of occurrence of such large-amplitude pulses.
While several authors have recently investigated the amplitude distributions of Crab giant pulses \citep{cbh04,ps07,bhatetal2008}, the most extensive work is by \citet{lundgrenetal1995} who analyzed 70 hr of observation from the Green Bank 43-m telescope at 812 MHz.\footnote{\citet{cm03} presents a scaled version of their results for a pulse width of 100 $\mu$s.} Unfortunately, the distribution is poorly constrained at such large amplitudes due to small number statistics. In any case, within the statistical errors, the rate at which such bright pulses may occur is rather low ($ {\rm < 1 ~ hr^{-1} }$). Counter to this is the possibility of pulsars that emit giant pulses brighter than that of the Crab, as it would be surprising if the Crab turns out to be the one that emits the brightest giant pulses in the local Universe. Indeed, giant pulses with amplitudes as large as 50--100 kJy have been detected at 1--2 GHz (or at least in the L band) \citep{ps07,bhatetal2008}, but these are relatively narrow ($\la$ 10 \us in duration) and so will be detected at a reduced sensitivity if the data are sampled at a rate of 50--100 \us, as is typically the case with pulsar search observations.  

\begin{figure}[t]
\epsscale{1.00}
\plotone{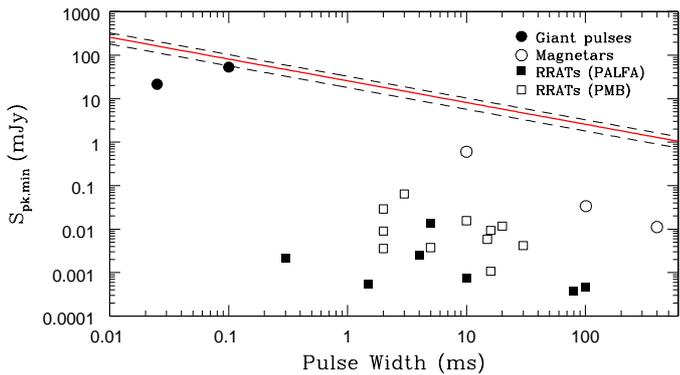}
\caption{Sensitivity curves for short-duration transient searches with the ALFA system, plotted as minimum detectable peak flux density (${\rm S_{pk,min}}$) when the matched filter width is equal to the pulse width. The solid curve is for a 5-$\sigma$ detection threshold and the sensitivity of the center beam; this curve will shift slightly upward or downward (i.e., dashed lines) for outer beams or for lower detection thresholds (e.g., 3-$\sigma$). Observing bandwidth assumed is  100 MHz and the center frequency is 1.4 GHz.  The plotted symbols denote transient sources known within our Galaxy (i.e., giant-pulse emitters, radio magnetars and rotating radio transients), after scaling their measured peak flux densities to M33's distance.} \label{trans}
\end{figure}

Rotating radio transients are a newly discovered class of 
objects with amplitudes ranging from 10 mJy to 4 Jy and pulse widths from 
1 to 100 ms \citep{mclaughlinetal2006,denevaetal2009,burke-bailes2010}. While the ALFA 
receiver is almost an order of magnitude more sensitive than the Parkes multibeam (PMB) receiver \citep{cordesetal2006}, the PALFA survey uses much shorter dwell times than the PMB survey (67--268 s for the former \vs 2100 s for the latter). This difference might explain a general tendency where the PALFA survey detections are relatively weaker and tend to occur more frequently.  The combination of ALFA's high sensitivity and our longer duration per pointing (2--3  hr) makes our search theoretically more sensitive to objects over a wider range of brightness and event rate. 

The achievable sensitivities for transient searches with ALFA (for a 5-$\sigma$ detection threshold) are shown in Fig.~\ref{trans}, along with the expected peak flux densities for currently known radio transients, if they are moved to a distance of $\sim$1 Mpc.  Evidently, the maximum achievable sensitivity is currently the fundamental limitation in finding such objects at M33's distance. With even the brightest known objects well below the detection threshold, prospects are not good unless there is a population of much brighter objects in the galaxy. 

Another interesting class is magnetar-like objects that may turn ``on'' at radio wavelengths during their active phases. 
These are young, long-period pulsars with ultra-strong surface magnetic fields ($\sim10^{14}$ G) whose emission is thought to be powered by the rearrangement and decay of their extreme magnetic fields \citep{duncan-thompson1992}.
They provide crucial links between normal pulsars and the class of anomalous X-ray pulsars (AXPs; \citet{mere2008}). 
Their radio fluxes tend to vary dramatically over timescales of days or weeks \citep{camiloetal2006} and thus they can potentially be detected through either periodicity searches or through searches of single pulses, depending on the brightness. 
With just three objects known so far (XTE J1810$-$197 and 1E 1547$-$5408 by \citet{camiloetal2006,camiloetal2007} and PSR J1622$-$4950 by \citet{levinetal2010}), their population remains poorly constrained. 
 However, if we take these three objects as representative of the class, their detection prospects are rather poor at M33's distance. The brightest of the three objects has $\Lum \la 0.15 \, \lumuu$, i.e., almost two orders of magnitude below the detection limit at a distance of $\sim$1 Mpc (see Fig.~\ref{sens}). The brightest pulse detected from XTE J1810$-$197 has $S_{\rm pk,1400}\la$ 40 Jy, \Wp $\la$ 10 ms, which is still an order of magnitude smaller than that required to enable a detection at $\sim$1 Mpc (Fig.~\ref{trans}). 
   
In addition to the above-described source classes, there are likely to be several as-yet-unknown types of sources that emit bursts of radio emission over short time durations. Among such hypothetical source classes are annihilating black holes and gravitational wave events \citep{rees1977,hansen-lyutikov2001}. The recent detection of a powerful burst (with a flux of 30 Jy and a width of 5 ms) by \citet{lorimeretal2007} triggered much interest in such exotic sources. The non-detection of more such signals in the PALFA survey argued against an astrophysical nature of its origin \citep{denevaetal2009} and more recently \citet{bsetal2010} convincingly described it to be of terrestrial origin. In any case, if any such bursts were to occur during our observations, they will be easily detectable. 

\begin{figure}[b]
\epsscale{1.00}
\plotone{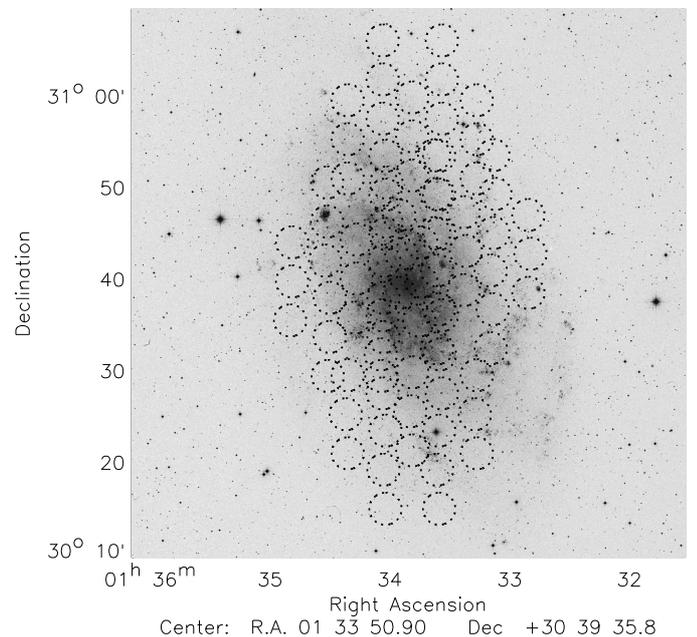}
\caption{ALFA pointings used for the M33 search overlaid on an optical image
of the galaxy from the Sloan Digital Sky Survey (York et al. 2000). The optical
radius of the galaxy is roughly 30 arcmin. There are 137 supernova remnant
candidates in the galaxy which are uniformly distributed out to a distance of
15 arcmin from the center (Long et al. 2010). As discussed in the text, the
optimal sampling possible with a maximum of 12 ALFA pointings still leaves a
small part of the galaxy sampled with a gain less than half-power of the on-axis
gain. The three closely spaced ALFA pointings, which fill in the area covered
by the ALFAÕs beam pattern as projected onto the sky (cf. Cordes et al. 2006
for details), are denoted as A, B, and C in Table 1. The galaxy coverage from
our 12 ALFA pointings almost fully overlaps with that of the 16 pointings of
McLaughlin \& Cordes (2003) used for their 430 MHz single-pixel search.}
\label{tiling}
\end{figure}

\section{Survey pointings and observation parameters} \label{s:survey}

Observations were made during September--December 2006 using Arecibo's ALFA system which allows simultaneous data collection from seven fields, each $\sim3.5^{\prime}$ (FWHM) across. Taking into account the hexagonal arrangement
of the beams on the sky and the near sidelobes, the combined power pattern is approximately $24^{\prime} \times 26^{\prime}$ \citep{cordesetal2006}. Thus, in order to cover a given field of the sky (with a sensitivity better than the half-power point of on-axis gain), 
three closely spaced ALFA pointings are needed in order to fill in the holes in the beam pattern projected onto the sky (see Fig. 3 of \citet{cordesetal2006}). 

Given M33's extent of roughly $73^{\prime} \times 45^{\prime}$, 12 such ALFA pointings are required to achieve a dense sampling of the galaxy. Details of pointings and observation parameters are summarized in Table~\ref{obs}. 
Fig.~\ref{tiling} shows the tiling scheme and pointings used for our survey. 
While this represents an optimal sampling possible with 12 ALFA pointings, it still leaves a small fraction of the galaxy unsampled. A complete sampling as per the ideal tiling scheme as used in the PALFA survey would require many more pointings, with a significant number of beams pointed away from the galaxy, and hence is not an efficient use of telescope time. We therefore optimize our search strategy by limiting the search to 12 pointings,
which cover approximately 95\% of the galaxy.  

For eight of these pointings, data were collected for 2 x 1 hr durations, while four other pointings were observed for 3 x 1 hr. Data were recorded using the Wideband Arecibo Pulsar Processor (WAPP) spectrometers \citep{dowdetal2000}, which were used to synthesize 256-lag autocorrelation functions for both three-level quantized polarization channels; correlation functions for the two channels were summed before writing out to disk. We observed a 100 MHz passband centered at 1440 MHz in each of the seven telescope beams; therefore, each frequency channel is approximately 390 kHz in width. 
In order to accommodate the constraints of the available data storage resources, approximately half of these data were sampled at a rate of once every 64 \us while the remaining half at a slightly slower sampling of once every 100 \us (see Table~\ref{obs}). 
This resulted in an aggregate data rate of  approximately 7 x 30 GB per every hour of observation, amounting to a total data volume of 6 TB from 15 observing sessions. 
These data were transported to the Swinburne supercomputing facility where the search processing was carried out. 

The RFI environment at the Arecibo Observatory is known to be a strong function of time of the day, with a tendency for minimal occupancy in time and frequency between midnight to dawn hours. Fortunately, all our observations were made during such relatively radio-quiet hours of the day when RFI from terrestrial sources is generally at a minimum.  As we highlight in a later section, overall data quality is found to be excellent, with only a small fraction ($\la 0.1 \%$) of the data affected by RFI. 
Moreover, the scheduling of our observing sessions were coordinated with those of the PALFA survey, usually centered at 19 hr LST (i.e., a few hours prior to our sessions). Successful detections of test pulsars at the beginning of our observing sessions and the discovery/confirmation observations of several new pulsars made from the PALFA survey data collected during this period thus readily provide important integrity checks of the full observing system and processing pipelines. 

\bigskip

\bigskip

\section{Searches for transient signals} \label{s:searches}

As outlined in \S 2, our search processing targeted two kinds of signals: (i) bursts of emission from giant-pulse emitters such as the Crab, or intermittent sources such as RRATs or magnetars, or one-off bursts from hitherto unknown sources (e.g., Lorimer et al.); (ii) periodic signals from highly luminous pulsars or active magnetars. For searching for transient signals of the first category, the procedures that we adopted are very similar to those described in \citet{denevaetal2009}, except that our search extends to a wider range in parameter space as the time scales we search for range from 64 \us to several seconds. In summary, raw data were dedispersed over a wide range in DM and each dedispersed time series is searched for individual pulses with two different time domain algorithms: (i) matched filtering, which is the most commonly used technique in single pulse searching, and (ii) time domain clustering, which employs an algorithm similar to friends-of-friends and thus sensitive to clustering of samples above threshold in time. 

In the matched filtering technique, a pulse template is effectively convolved with the dedispersed time series, where a template is chosen to reflect the diversity of pulse shapes that are searched. For example, unscattered pulse shapes can be modeled as either a single or a sum of two or more Gaussians whereas asymmetric pulse templates would be a better approximation for scattered pulses which tend to have exponential tails.  In practice, we approximate matched filtering by progressive smoothing of the time series by adding up to $2^n$ neighboring samples and selecting events after a set threshold after each iteration. In light of the marginal detection prospects anticipated in our searches, we use a fairly low threshold of 3.5 $\sigma$ for event detection. 
As the smoothing is done in pairs of samples, the pulse template used is effectively a boxcar of length $2^n$ samples. 

The friends-of-friends method relies on a completely different approach, where the dedispersed time series is processed sequentially and if an event above the threshold is found it is designated as the first of a cluster. A cluster of events is then augmented while successive samples are found to be above the threshold. The brightest sample of a cluster is recorded as the event amplitude and the total number of samples in the cluster as its width.  
Such an approach has a maximum effective time it is sensitive to, which is related to
the block size of the data used in calculating statistics of the time series.
In our case, it is limited to 4096 time samples while searching for short-duration events (\la 1 s); i.e., 262--410 ms depending on whether the sample time used in the observation is 64 or 100 \us.

\begin{figure*}[t]
\vskip -0.05in
\epsscale{2.1}
\plottwo{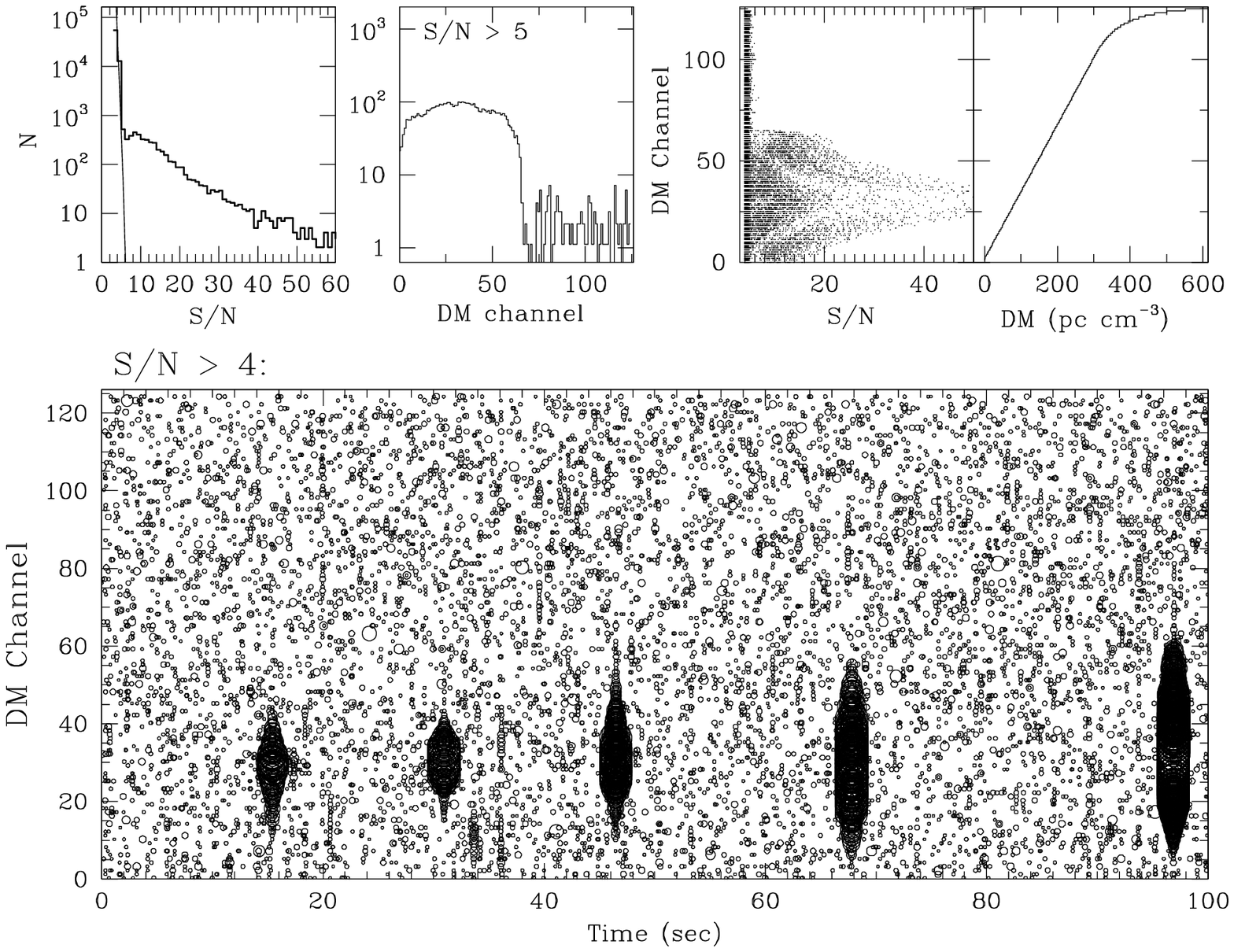}{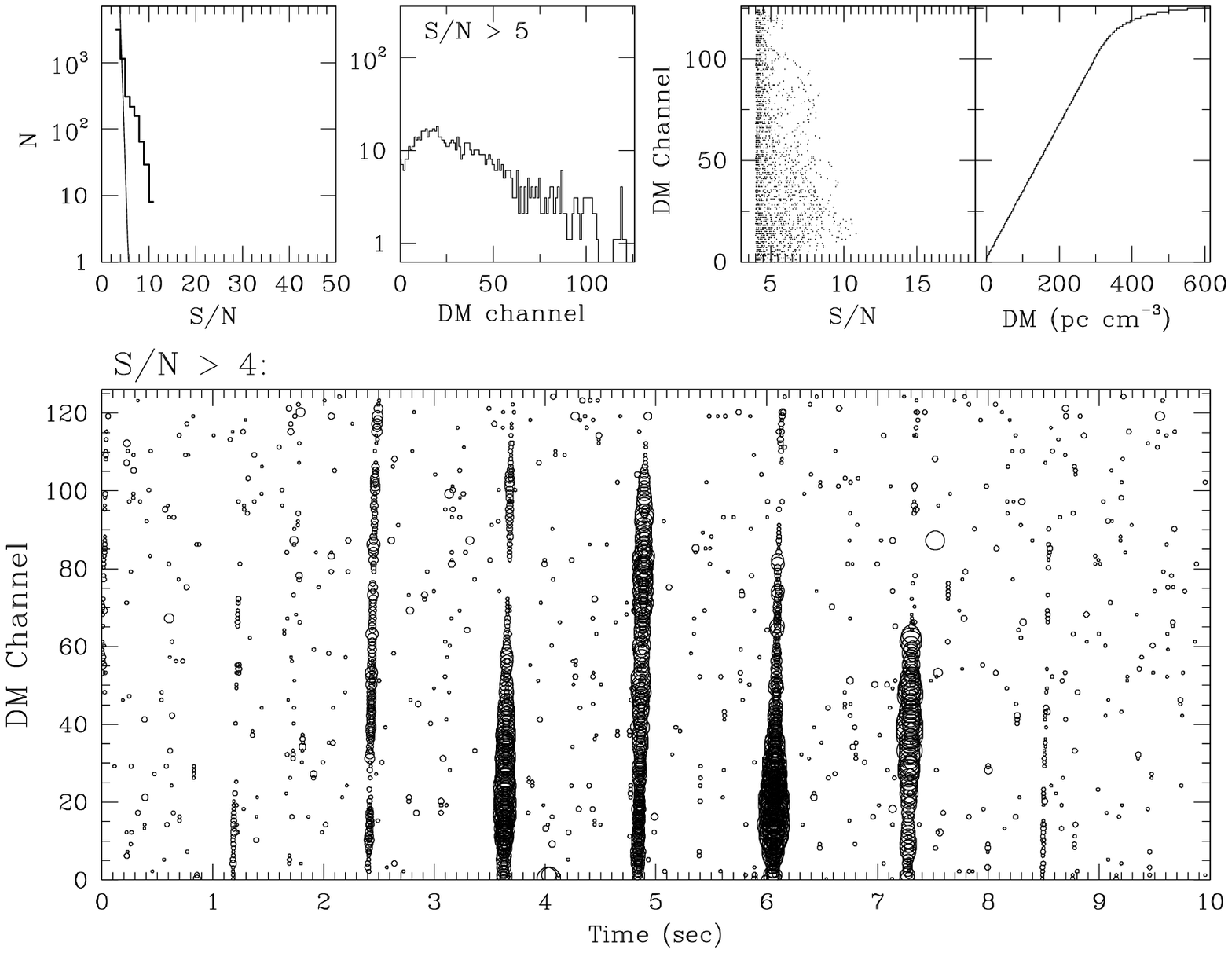}
\caption{Diagnostic plots from transient search processing: ({\it Left}) Detections of individual pulses from J0628+0909, a RRAT-like object discovered in the PALFA survey \citep{cordesetal2006}; ({\it Right}) Detections of single pulses from a known pulsar PSR B0045+33 that was routinely observed as a test pulsar at the start of every observing session. In each set, the top panels (left to right) show the statistics of events above 5-$\sigma$ against S/N and DM, the scatter plot of S/Ns and DMs, and the mapping from the DM channel to DM value. The bottom panel shows the summary of events (S/N $>$ 4) in the time-DM plane.  Multiple bright pulses are seen at the $\approx$1.2 s periodicity of the pulsar B0045+33; the brighter pulses tend to peak near the pulsar's true DM $\approx$ 40 \dmu (DM channel = 15). As the pulse width is 3 $\times$ larger than the dispersion sweep across the 100 MHz band, most pulses tend to appear at a large range in DM. The signal strength tends to peak near the true DM of the object, $\approx$ 88 \dmu, for J0628+0909. The size of the open circle is proportional to S/N.}\label{example}
\end{figure*}

\subsection{Data processing} \label{s:data}

We dedisperse the raw data over a wide range of 0--1000 \dmu in DM using 1272 trial values that were optimally chosen to span this range. 
Although the DM to M33 is estimated to be only $\sim$50~pc~cm${}^{-3}$, we set an upper limit of~1000~pc~cm${}^{-3}$ in order to account for any pulsars that might be in or behind \ion{H}{2} regions within M33 and to account for any potential extragalactic pulses originating from behind M33.
For searches based on the matched filtering scheme, we used boxcars of widths $2^n$, where $n$ ranging from 0 to 17, thus covering a large range in timescale from 64 us to 8.4 s. Even though the detected emission from fast transients known so far have timescales $\la$ 100 ms, this extended range allows a much more comprehensive search, as it targets even unusually broad pulses or strongly scattered pulse shapes. This strategy based on the use of symmetric templates is however non-optimal for scattered pulses (which are asymmetric due to their exponential tails), which will therefore be detected with a reduced sensitivity.

For processing efficiency, the raw data were broken into short blocks of 135 seconds each, which is set by the 2 GB file size limitation of the data recording. Moreover, the matched filtering procedure was done in two passes: the first pass with $n$ ranging from 0 to 7 (i.e., pulse widths ranging from 64/100 \us
 to 8.192/12.8 ms depending on the time sampling), and the second from $n$=8 to 17 (i.e., pulse widths ranging from $\sim$10 ms to $\sim$10 s), thus spanning almost five orders of magnitude in timescales. The combination of short data blocks and two separate passes for short and long events  ($\la$ 10 ms and $\ga$ 10 ms, respectively) avoids overcrowded diagnostic plots and thus a much more efficient process of sifting through the search outputs.  

As highlighted by \citet{denevaetal2009}, the time-domain clustering method is complementary to the matched filtering technique, with the major advantage being its ability to efficiently suppress many false positives that arise from RFI sources. This is because a broad RFI burst will most likely be detected as a single event by this method as opposed to a large number of events as is the case with matched filter method. A limiting factor however is that the sensitivity is limited by the sampling resolution as there is no progressive smoothing of the time series and therefore is it less sensitive to weak, narrow pulses. Any genuine signals of astrophysical origin in the data will most likely be detected by both the methods, albeit at different sensitivities depending on the pulse width. This is the case in the PALFA survey.

\subsection{Analysis of Search Outputs} \label{s:analysis}

Example diagnostic plots from our search processing are shown in Fig.~\ref{example}. These show all candidate events with signal-to-noise ratio (S/N) larger than the 5-$\sigma$ threshold and their statistics in the form of distributions of S/N and DM as well as a scatter plot of S/N \vs DM. Any genuine signals of astronomical nature will have distinct signatures on such plots; for instance, they will be detected over a contiguous range of DMs, peaking near the true DM and with a signal strength that falls off smoothly with an increasing departure from the true DM. Ideally, in the absence of RFI, there will be a clear peak in the DM distribution as
well as in the scatter plot (S/N against DM), peaking near the true DM of the signal.
The width of the peak \vs DM is broader for broader pulses.  
The density of events on the scatter plot will depend on the number of similar events detected in the data. 
In contrast, RFI with narrow time structure will peak at zero DM and 
fall off rapidly  at larger DMs unless it has intrinsic swept-frequency
structure.  If very strong or temporally wide,  
the RFI can appear over the full 
range of DMs. Such strong and impulsive RFI 
bursts can potentially modify or even mask the signatures of 
any real signals present in the data. 

\begin{figure*}[t]
\epsscale{1.50}
\plotone{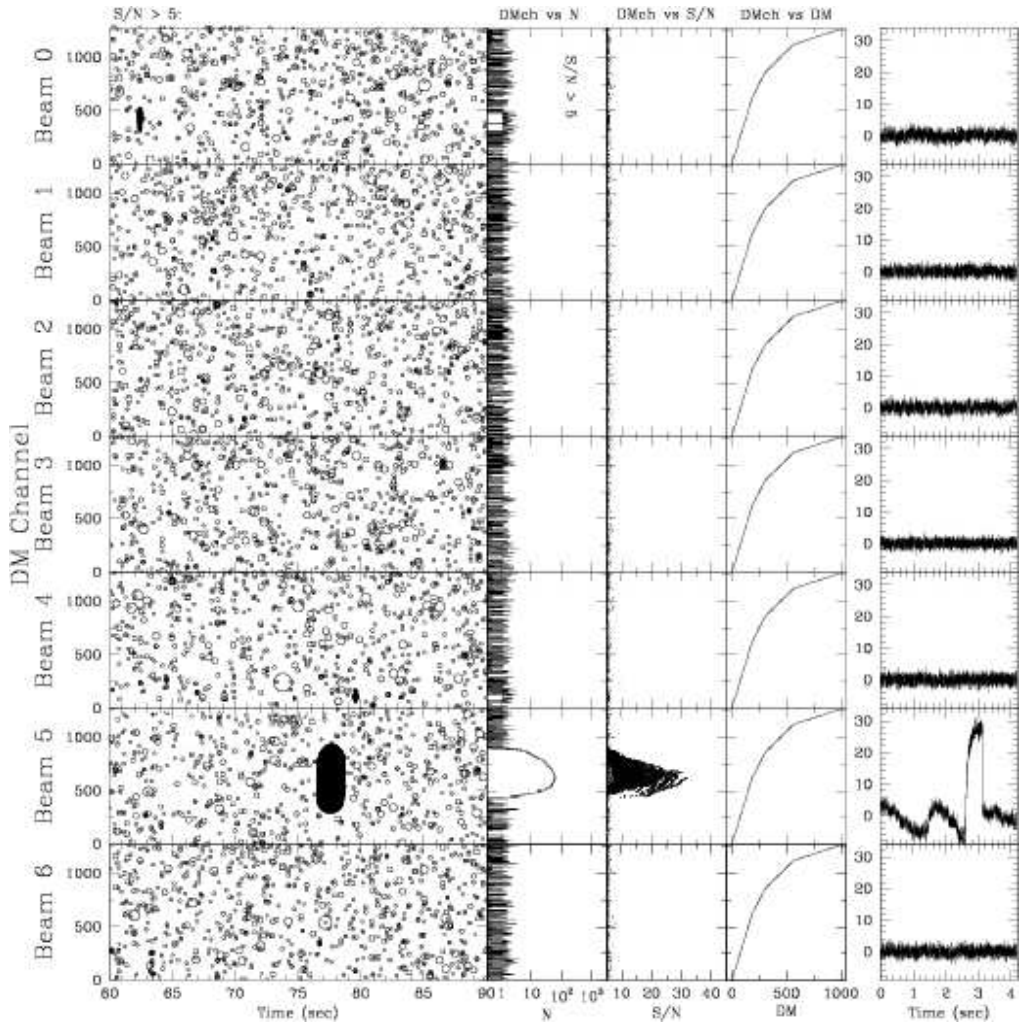}
\caption{ Example diagnostic plots from event analysis and classification: An isolated event from observations made on MJD 53992 toward the pointing M33.4.C,  detected with a signal strength (S/N $\sim$30) that peaks near DM$\sim$200 \dmu in beam 5, indicating a broad, dispersed pulse. A closer scrutiny reveals this tantalizing event to be of instrumental origin, largely a manifestation of an abrupt increase in power level that lasted for a short duration. The localization in DM is intriguing and an explanation is given in \S~4.4. Each row shows results from one ALFA beam along with  dedispersed time series. From left to right, panels show: events with S/N $>$ 5 \vs DM channel and time, number of events \vs DM channel, event S/N \vs DM channel, mapping from the DM channel to DM value, and time series of a short data segment around the event.}\label{allbeam} 
\end{figure*}

In principle, a careful assessment of these diagnostic plots should allow identification of genuine signals \vs those arising from RFI; however, the efficiency of such a strategy critically depends on the nature and strength of RFI.  As is well known, impulsive RFI can often mimic the signatures of real signals and may manifest as islands in the time-DM plot or as distributions which peak at certain DMs. Indeed, as we discuss in \S~\ref{s:event}, a more rigorous scrutiny that involves inspection of relevant raw data segments in the time-frequency plane in order to confirm the expected dispersion sweep will help detect and eliminate all such cases with high enough signal-to-noise for frequency-time analysis. However, the large number of events typically detected in our searches make such an approach both tedious and impractical. 

Fortunately, multiple beams of the ALFA system can be well exploited for this purpose given the significant improvement it offers for event analysis. However, such a filtering scheme based on multiple beams is not straightforward because scattering off telescope structures can introduce RFI into the telescope optics.  
Depending on the telescope orientation, RFI bursts may be detected in all, some or none of the ALFA beams. We therefore adopt a scheme whereby search outputs from multiple beams are compared and inspected visually, as described in the next section.  

\begin{figure*}[t]
\epsscale{1.5}
\plotone{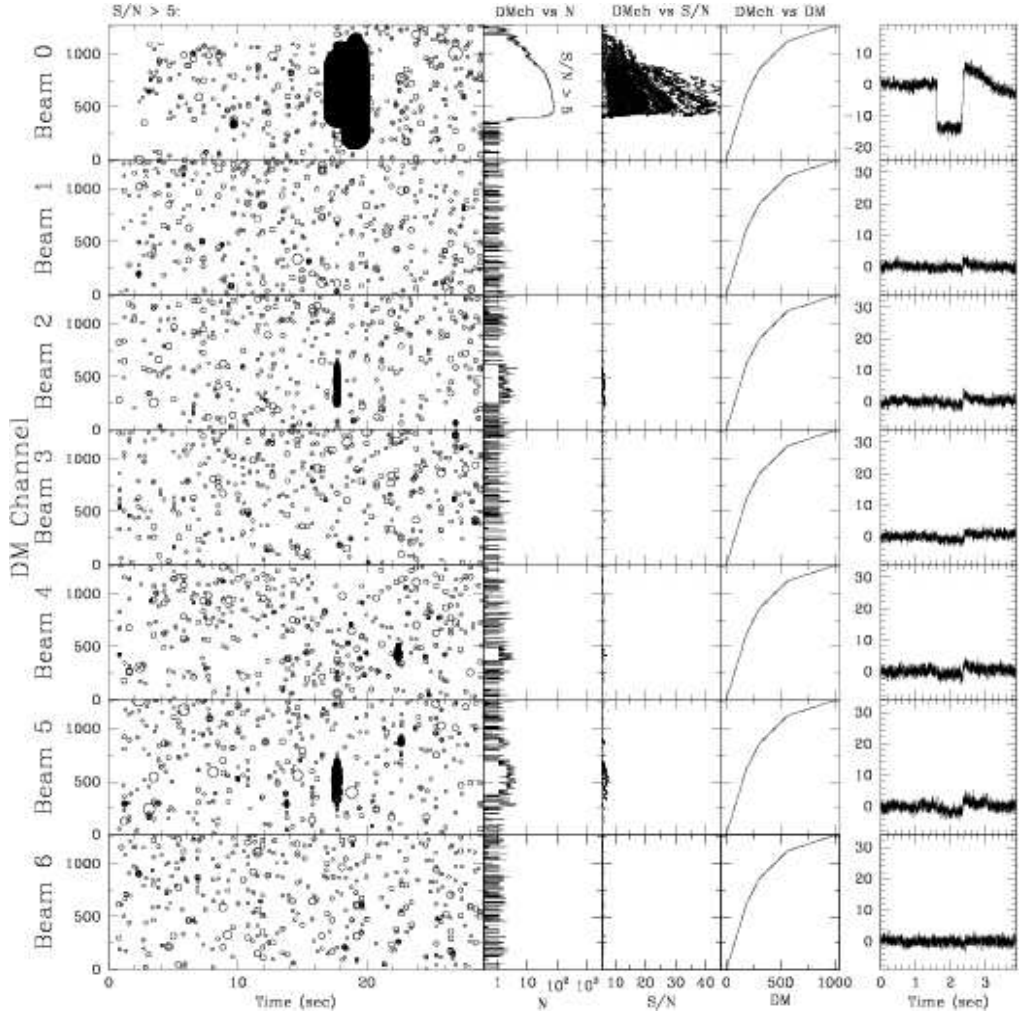}
\caption{ Example diagnostic plots from event analysis and classification: a spurious event from observations made on MJD 53995 toward the pointing M33.2.B, which appears at a range of DMs and signal strengths simultaneously through the center beam and the beams 2 and 5. The event was classified to be of RFI origin, given a  highly asymmetric nature of the S/N \vs DM curve (in beam 0), and its appearance (with much lower S/Ns) in two well-separated outer beams (2 and 5). Each row shows results from one ALFA beam along with  dedispersed time series. From left to right, panels show: events with S/N $>$ 5 \vs DM channel and time, number of events \vs DM channel, event S/N \vs DM channel, mapping from the DM channel to DM value, and time series of a short data segment around the event.}\label{allbeamtwo}
\end{figure*}

\subsection{Event classification and Scrutiny} \label{s:event}

In order to aid a meaningful classification of events and their final scrutiny, we assemble various search output plots as described in \S~\ref{s:analysis} for all 7 beams to generate multipanel plots as shown in Figs.~\ref{allbeam}~and~\ref{allbeamtwo}. Such plots readily provide a comprehensive summary of all detected events in a given data set as well as data quality, and thus drastically improve the efficiency of event analysis. Several thousands of plots generated in this manner are included in a MySQL database which was later accessed and queried using a python-based interactive viewer for ranking and classification. We adopted a scheme whereby each data segment is first assessed and ranked, with ranks of 1 and 2 (Class I and II) to reflect prospective candidates that require further careful scrutiny, while data that are devoid of any significant events (i.e., no clustering in the time-DM plane or in the statistics plots) were given a rank of 3 and those which identified to contain {\it at least} one event of RFI origin were given a rank 4. The best candidates are thus of Class I, while those of Class II are considered promising. Results of this assessment are summarized in Table~\ref{stats}. The RFI bursts are found to typically last over time durations of a few seconds or less, and no more than a few instances detected in a given data segment. Thus only a small fraction of the data are influenced by RFI and therefore overall data quality is pleasingly good.  

Candidate events of Class I and II are then subjected to a more rigorous scrutiny, where short raw data segments (4 seconds) centered at the time of the event were extracted for all seven beams and displayed in the frequency-time plane (i.e., dynamic spectrum) in order to check the expected dispersion sweep across the frequency band. This provides a crucial diagnostic for celestial \vs terrestrial origin. These dynamic spectra plots are examined along with the dedispersed time series for the event's DM, to facilitate a more comprehensive scrutiny of the candidate event. For the beam corresponding to the candidate event, further scrutiny is carried out by also inspecting the dedispersed time series at two nearby DMs and at zero DM. This analysis is typically done at the original resolution of the data (64 or 100 \us). However, for candidate events with $S/N \la 10$, and thus too weak to be discernible in the frequency-time plane, data are suitably smoothed in time and/or frequency and then re-inspected. Such a procedure allows a rigorous scrutiny and the final assessment is done in this manner for all 169 candidate events that emerged from our search. 

Example plots illustrating this scheme are shown in Fig.~\ref{quickg}, which shows the signatures of a real signal (an individual pulse detected from J0628+0909, a RRAT-like object that was originally discovered in the PALFA survey \citep{cordesetal2006}) and an RFI event.

In summary, our scheme for event analysis, which relies on multiple diagnostic plots, detailed comparisons with results from other beams and careful examinations of the raw data segments, have allowed us to conduct a thorough and comprehensive assessment of all prospective candidate events. Unfortunately, we have not found any signal for which an astrophysical origin can be convincingly attributed. Instead, all such events (Class I and II) turned out to be intriguing manifestations of different kinds of RFI, with the pulse widths ranging from $\sim$100 \us to seconds and DMs from $\sim$10 to $\sim$300 \dmu. The pulse shapes vary from highly spiky to rectangular, and from smoothly symmetric to highly exponential, or sometimes seen as trains of short bursts. More details on the nature of RFI and some quantitative analysis are described in \S~\ref{s:radio}. 

\subsection{Radio Frequency Interference and Data Quality} \label{s:radio}

Spurious events that originate from impulsive RFI are a major problem in transient detection. However, as highlighted in previous sections, the overall data quality is found to be pleasingly good in our observations and only a small fraction  of the data ($\la0.1\%$) are corrupted by RFI.  
Moreover, our extensive scheme of event scrutiny that relies on multiple diagnostics including the event signatures in the time-DM plane, signal strength \vs DM and its appearance across multiple beams, gives us a powerful handle in the identification and elimination of the bulk of such RFI-generated events. 
In particular, the infamous, and often periodic, strong RFI pulses from the Federal Aviation Administration (FAA) radar at San Juan airport, which have a 12-second periodicity and are $\sim$1 second wide (see \citet{denevaetal2009} for details), are practically absent in our data. 
This is most likely to do with the fact that our observations were carried out during
the radio-quiet hours between midnight and dawn, when the FAA radar tends to be
relatively less active.
Furthermore, in the small number of instances when such radar pulses were seen, their nature is found to be rather aperiodic. Given such low presence of radar RFI,
a rigorous excision scheme as discussed by \citet{denevaetal2009} is not warranted for our data. 



\begin{figure*}
\begin{center}
\includegraphics[angle=-90,width=0.70\textwidth]{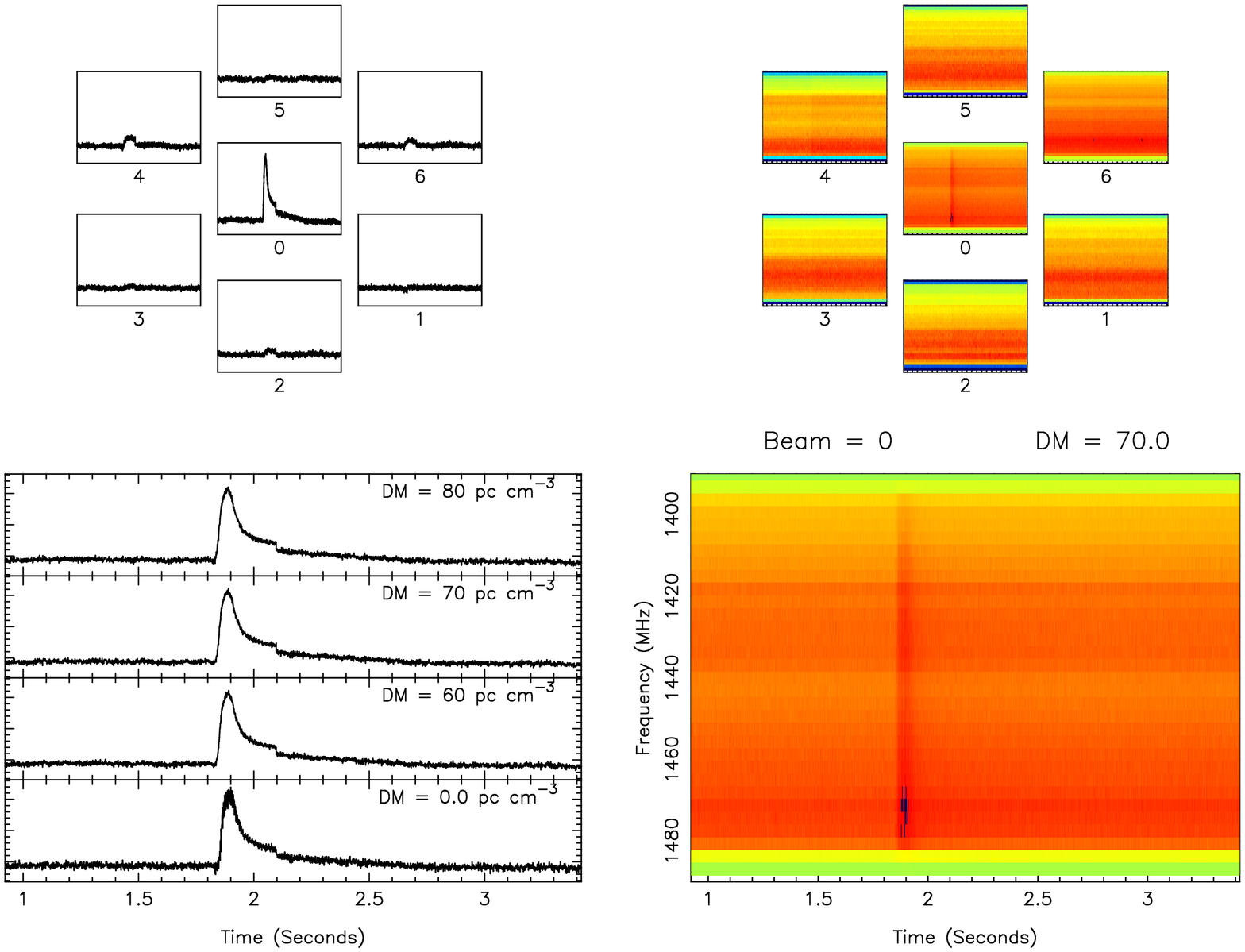}
\vskip 0.5in
\includegraphics[angle=-90,width=0.70\textwidth]{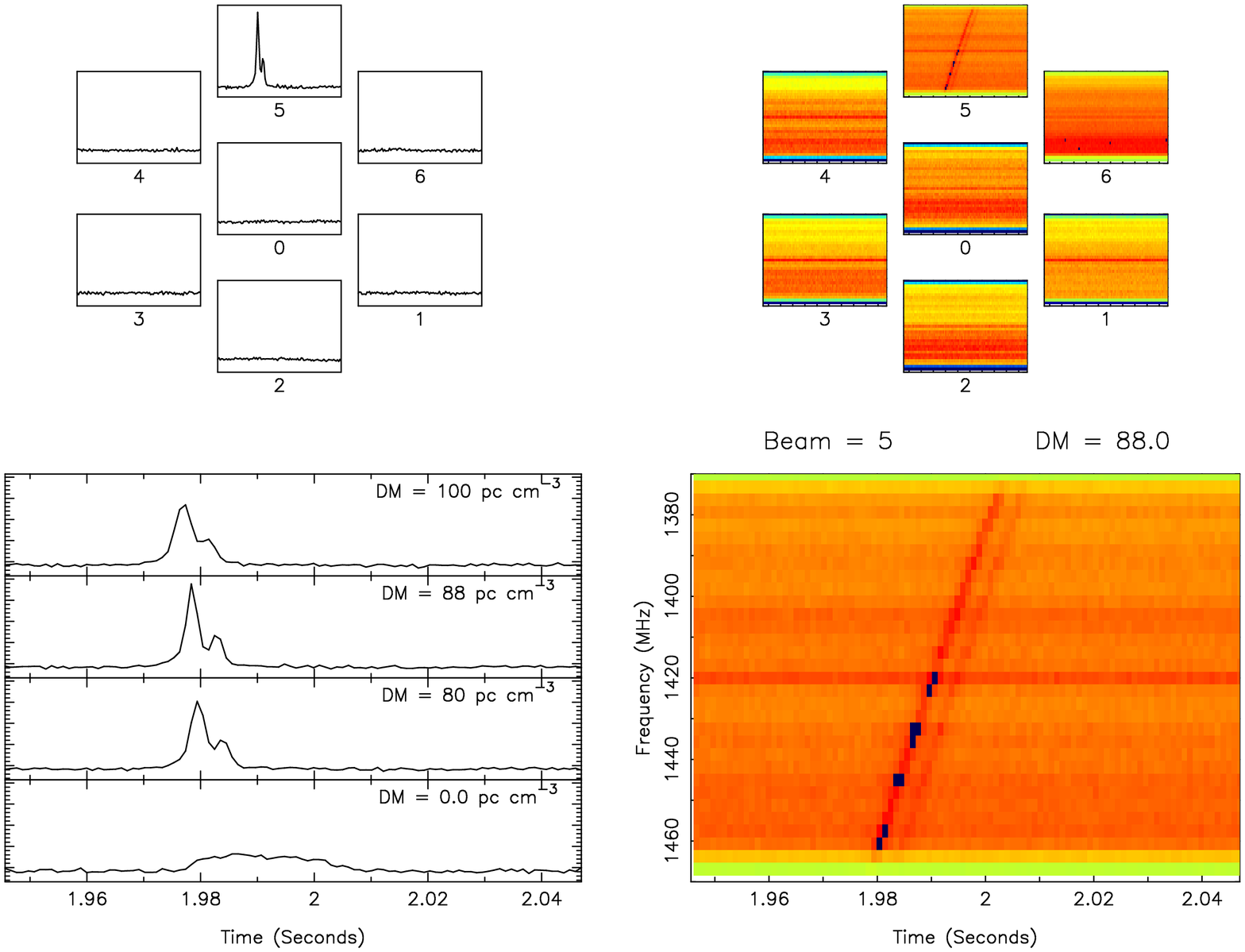}
\end{center}
\caption{
Example diagnostic plots from event analysis: the top two rows show a spurious event at DM$\sim$70 \dmu produced by impulsive RFI detected through the center beam of ALFA, while the bottom two rows show a real signal from J0628+0909 discovered in the PALFA survey \citep{cordesetal2006}. In each set, the top rows of multiple panels show the dedispersed time series and dynamic spectra plots for all 7 beams, and the bottom panel additional diagnostic plots for the beam in which 
the candidate event was detected. See the text (\S~\ref{s:event}) for further description. The time-frequency plot of the J0628+0909 pulse clearly displays frequency structure that is caused by interstellar scintillation.}\label{quickg}
\end{figure*}

Table~2 provides a short summary of the number of candidate events identified in our analysis and those which can be readily identified as RFI. Our assessment of an event to be of RFI origin  is based on one or more criteria, such as a high signal strength at zero DM and its appearance in many or all DMs, as well as in several or all 7 beams. As seen from this summary, the total number of RFI events detected depends on the detection method, with the time domain clustering method comparatively more efficient in their detections. On the other hand, the matched filtering technique results in a much larger number (four times) of ``interesting'' events. In general, with any broad bursts (RFI or real), we would expect the matched filtering technique to result in a larger number of events  than the time domain clustering method (cf. \S~\ref{s:data}), however the number of events will depend on factors such as the pulse width, block length and the number of boxcars used in processing. As discussed in \S~\ref{s:event}, all such promising candidates are subjected to a thorough scrutiny, involving the extraction of the relevant raw data segments and their inspection both in the time-frequency plane as well as in time series at multiple DMs, across all seven beams (Fig.~\ref{quickg}).

Figures~\ref{allbeam}~and~\ref{allbeamtwo} show representative examples of such interesting candidate events seen in our analysis. In many cases, the most striking feature is the detection over a contiguous range of DMs, with the signal strength peaking at a certain DM, and its occurrence in only one or a few beams. While at first glance such signatures may suggest a likely non-terrestrial origin, our close scrutiny reveals that they are  often manifestations of intriguing RFI pulses. 
For instance, broad, asymmetric pulses such as those in Fig.~\ref{quickg}, with a significant rise time ($\sim$100 ms) and a pulse width ($\sim$250 ms) that is comparable to the block length used in our analysis, can potentially mimic a localization in the DM space, even though it is intrinsically at DM=0. 
The same may apply to jumps in power levels that last over durations of the order of block length or longer (e.g., Fig.~\ref{allbeam}). 
The localization in DM may occur when the pulse width (or the duration of the event) is comparable to the block length used in processing and if at least one of the pulse edges occurs near the edge of the block.
In such cases, de-dispersion results in smearing of the burst and, for certain DM values, the burst may be split into two successive blocks after de-dispersion. 
At both low and high DMs, much of the burst will still lie within a
single  block, which means larger apparent detection thresholds.
Consequently, the burst may not be detected as an event.
The localization in the DM space that arise in this manner is thus purely a processing artifact.  This is especially the case in our processing pass targeted at ``short" events (time durations $\la$10 ms) where the block length used in determining the noise statistics is restricted to 4096 time samples ($\sim$400 ms), which is comparable to maximal dispersion sweep (278 ms at DM=1000 \dmu). 
The extent of DM localization will primarily depend on the rise time of the pulse and, to a certain extent, other signal characteristics such as the pulse amplitude and width, and in general longer the rise time broader the extent of localization. 
We have inspected all such candidate events detected in our analysis and found none that may warrant attributing an astrophysical origin. In many cases they were recognized as RFI pulses from the FAA radar (sporadic or aperiodic in nature), and in some cases sudden jumps in intensity levels that are likely of instrumental origin. 


\section{Searches for periodic signals} \label{s:periodic}

Data were also searched for periodic signals, for which we used the {\tt PRESTO} search code\footnote{http://www.cv.nrao.edu/$\sim$sransom/presto}, following procedures similar to those adopted for the offline processing of the PALFA survey data. 
In summary, the raw data were first dedispersed over a large range of DM, 0--1000 \dmu, subdivided into four groups, with successive ranges using a larger step size and coarser time resolution. 
Specifically, a total of 1272 trial DMs were used for the search: with 624, 240, 264 and 144 trial DMs spanning the 0--1000 \dmu range in steps of 0.3, 0.5, 1.0 and 3.0 \dmu respectively.  For the last three DM groups, the data were progressively smoothed and decimated (in time)  by a factor 2, 4 and 8 respectively. 

A Fast Fourier Transform was performed on the resulting dedispersed time series and the power spectrum was searched for periodic signals by summing up to 16 harmonics and applying a harmonic sum threshold of 6$\sigma$. With our 1 hr long integration ($ > 2^{25}$ samples) for each pointing, this translates to processing 
7 x 60 GB data per pointing, typically producing hundreds
of candidates per each beam. The large number of candidates that emerged from this search (11,879 in total) were loaded into a MySQL database and subjected to 
a careful inspection for any prospective candidate using an interactive viewer tool. 

\begin{figure}[t]
\epsscale{1.00}
\plotone{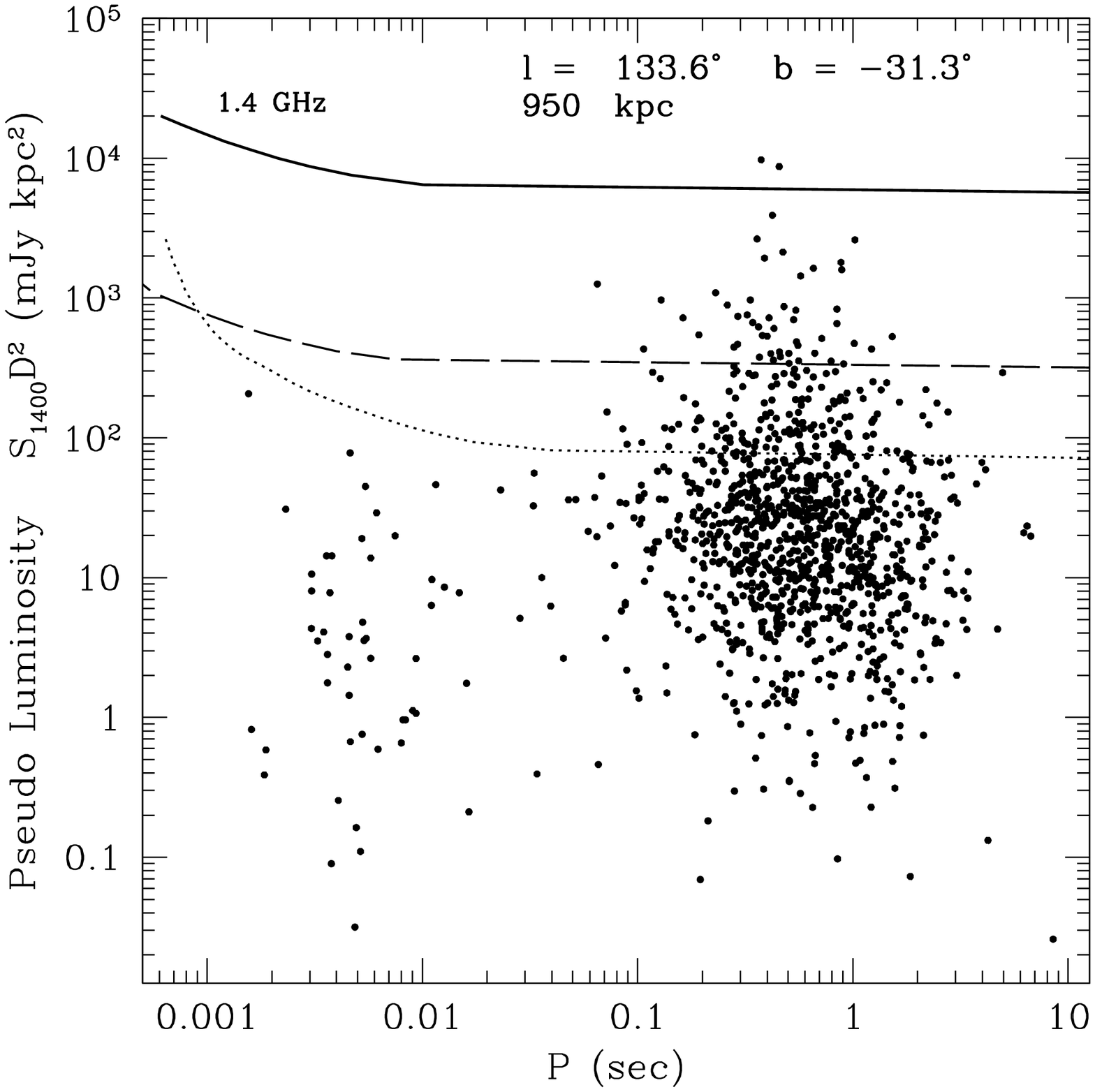}
\caption{Plot of pseudo-luminosity $S_{1400}D^2$ against the spin period, assuming M33 harbors 
a pulsar population that is very similar to that of the Milky Way. The data points are from the ATNF
pulsar catalog \citep{manchesteretal2005}, scaled for a distance of 0.95~Mpc, where M33 is located. 
The solid curve is the $6\sigma$ detection threshold at 1.4~GHz based on 
the observing parameters described in the text and for identification of 
a preliminary set of periodicity candidates. 
The other two curves indicate the detection threshold for a survey 
with the Square Kilometer array for a 1-hr integration at
2~GHz with a 1~GHz bandwidth (dashed line) and at 0.43~GHz with a 
0.1~GHz bandwidth (dotted line).  
A sensitivity $A/T = 10^4$~m$^2$~K$^{-1}$ is assumed for the SKA. 
\label{fig:lp_vs_p}
}
\end{figure}

A thorough scrutiny of these candidates revealed $\sim$10\% to be above an 8-$\sigma$ detection threshold as discussed in \S~\ref{s:periodicity}. Not surprisingly, a large fraction of them turned out to be of spurious origin, essentially manifestations of strong RFI pulses originating from the FAA radar (see \S~\ref{s:radio}). Such candidates are generally detected with a roughly 6-s periodicity or its harmonics (i.e., P $\sim$ 3, 2 or 1 s) at low or moderate DMs (\la 50 \dmu), and inspection of the relevant time series show multiple instances of such RFI bursts and a folded pulse profile that has a characteristic asymmetric (exponential) shape. Only a tiny fraction ($\sim$2\%) of the low S/N candidates (6--8 $\sigma$) was found to be caused by the radar RFI.

Of the remaining candidates, roughly 40\% were found to be within a narrow region of the parameter space (DM \la 5 \dmu, P \la 5 ms), while the majority ($\sim$60\%) were weak, short-period candidates spanning a large range in DM. We inspected the diagnostic plots for all candidates at DM $>$ 0 and above 6-$\sigma$, and found none that satisfied the standard assessment criteria to be considered real (e.g., persistence of signal \vs time and \vs frequency, a clear peak in the signal strength \vs DM and in the period-period derivative search space, etc.). As seen from Fig.~\ref{sens}, our integration times of 1--2 hr will translate to limiting flux densities of $\sim$3--10 \uJy at M33's distance, depending on the duty cycle (\Wp/P $\sim$1--10\%). 

\section{Implications for source luminosities and populations} \label{s:impli}


Figure~\ref{fig:lp_vs_p} shows the sensitivity curve for our analysis.
The solid curve takes into account radiometer noise, and 
propagation effects in the Milky Way (dispersion and scattering) using
the NE2001 model \citep{ne2001}.
At the Galactic latitude of M33 ($b \approx -31^{\circ}$) and for our observation frequency (1.4 GHz),
pulse broadening from residual dispersion smearing and from scattering
in the Milky Way or from M33 are negligible.    
For simplicity, we have assumed that M33 harbors a pulsar population 
similar to that in the Milky Way. 
Only a few known
pulsars are just above the detection curve, implying that if there
were $\sim$2000 or more active pulsars that beam toward us from
M33, we should have detected a few of them.
However, the mass of 
M33 is about 10\% of the Milky Way's \citep[e.g.,][]{sakamotoetal2003}, so there may be
fewer than 2000 detectable pulsars in M33.  
A counter argument may be in order if the expected neutron star population in M33 is larger. 
For instance, while we may expect the star formation rate in M33 to be roughly 10\% of that in the Milky Way 
based on mass arguments, the measured rate is about 10--25\% \citep{fermi2010}, suggesting that there may be 
more neutron stars than we may expect based on the 10\% rate.

The figure also gives curves for the Square Kilometer Array at frequencies 
of 0.43 and 2~GHz.   We have assumed a sensitivity of 0.28~Jy, corresponding 
to an effective area divided by system temperature of $10^4$~m$^2$~K$^{-1}$.  
The nominal sensitivity for SKA Phase 1 is approximately the same as that of 
Arecibo \citep{garretetal2010}, while the full SKA may have a sensitivity that is 
several times to a factor of ten larger than that of Arecibo. 
 
A simple upper bound can be placed on transient sources in M33.
We estimate the volume sampled per telescope beam diameter 
$\theta_{\rm b} \approx 3.5$~arc~min as
$\Delta V_1 = (\pi/4)(D_{\rm M33}\theta_{\rm b})^2 H_{\rm eff}
	\approx 0.08(H_{\rm eff}/0.1~{\rm kpc})$~kpc$^3$,
	where $H_{\rm eff}$ is the extent of the galaxy sampled 
	along the beam direction (assuming a face-on orientation of
	the galaxy).   
With a total aggregate search time of 30~hr (c.f. Table 1), the limit
on the event rate per unit volume is
$\dot n_s \la \left(7\Delta V_1 T_{\rm total}\right)^{-1}
	\approx 0.06\left(0.1~{\rm kpc} / H_{\rm eff} \right)
	{\rm events~kpc^{-3}~hr^{-1}}$ 
for events that satisfy
$\Spk \ga 23.5~{\rm mJy~W_{\rm ms}^{-1/2}}$.
The Crab pulsar emits pulses with amplitudes consistent with
this expression (when instrumentally broadened to 100$~\mu s$ and accounting for the ratio of
distances of the Crab pulsar and M33) about once per 10~hr, 
so the event rate limit corresponds to an upper bound of
about one Crab-pulsar-like object per kpc$^{3}$ in M33. 

It is well known that pulsars have high space velocities with
a distribution that extends well beyond the local Galactic escape 
velocity $\sim 500$~km~s$^{-1}$ \citep{arzoumanianetal2002}.   
For the 84 beam areas of our
survey, the volume out to a distance of 5~kpc (i.e., the distance 
traversed by a canonical pulsar moving at 500~km~s$^{-1}$ 
in its 10 Myr lifetime) above the Galactic disk is $\sim 3$~kpc$^3$.  
Assuming roughly 50\% of the known 
$\sim2\times10^3$ active radio pulsars that beam toward us are 
on escape trajectories, there should be about three pulsars in the 
search volume. The lack of any detections is consistent with this 
estimate given that most of the pulsars will not reach 5~kpc during 
their radio-emitting lifetimes. 

Lorimer et al (2007) reported a single 30-Jy, 5-ms event from their survey of the Magellanic Clouds that covered $\sim$9 ${\rm deg^2}$ at high Galactic latitudes.  
Assuming a source population that is homogeneously distributed in Euclidean space, Deneva et al. (2009) estimate $\sim$600 such events above the detection threshold of the PALFA survey. 
These limits follow from the calculations where the number of detected events $ N \propto  \Omega \, T_{\rm obs} \, \Dmax $ (where $\Omega$ is the solid angle coverage of the beam and $T_{\rm obs}$
 is the survey duration) and by ignoring the propagation effects that may limit the detectability of bursts from distant sources. 
 The maximum distance to which an event of a given peak flux density $\Spk$ and the pulse width W is detectable is given by $ \Dmax \propto ( m \, \Ssys )^{3/2} \delnu ^{-3/4} $ (cf. Equation~\ref{eq:trans}, where $m$ is the detection threshold in units of $\sigma$). 
The sensitivities of our single-pulse searches are similar to those of the PALFA survey,  however our total survey duration is substantially smaller (30 hr \vs 461 hr). 
We thus estimate $\sim$40 pulses above our detection threshold of  $\sim$20 mJy (cf. \S~\ref{s:transient}). 
In any case, the non-detection of such plausible extragalactic events in our survey reinforces the conclusion of Deneva et al., effectively precluding a source population that has a   homogenous 
and isotropic  distribution in the sky. 

\begin{figure}[b]
\begin{center}
\includegraphics[angle=-90,width=0.45\textwidth]{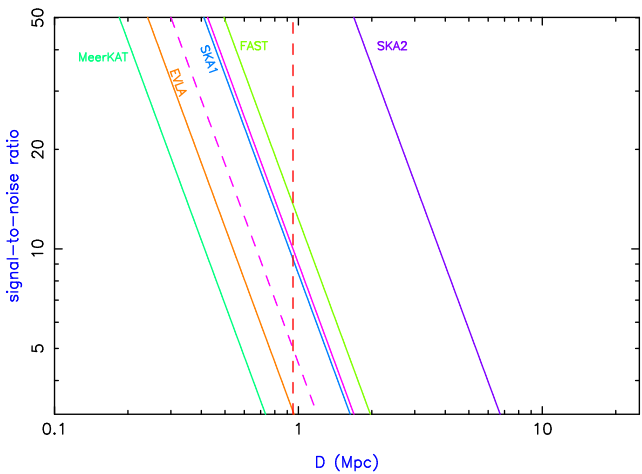}
\end{center}
\caption{Distance to which a bright giant pulse --- having a flux density of 20 kJy,
from the Crab pulsar (2 kpc) --- could be detected with future instruments. See
also Equation (2) and Table 3. The vertical dashed line indicates the distance
of M33. For clarity, the curve for Arecibo is not labeled, but it is the purple
curve between the curves for FAST and SKA-1. The dashed purple curve is
for our current Arecibo search. We do not plot above a signal-to-noise ratio of
50, though it is clear that values exceeding this limit are possible for different
combinations of pulse flux densities and telescopes.}
\label{fig:future}
\end{figure}

\section{Future Prospects} \label{s:future}

Detection of extragalactic pulsars and transients is inevitable with
continued observations with existing telescopes and observations with
forthcoming instruments.  As discussed in~\S\ref{s:intro}, the
\objectname{Crab pulsar}'s giant pulse amplitude distribution is a
power law, with no upper limit having yet been detected.  Further, the
rate of giant pulses from the \objectname{Crab pulsar} is also a power
law in pulse energy, with brighter pulses being less frequent than
fainter pulses.  However, because no upper limit to the pulse
amplitude distribution has been observed, longer dwell times imply a
higher probability of brighter, and therefore more detectable, pulses.
Further, it is plausible that objects brighter than the
\objectname{Crab pulsar} exist.

There are now a suite of centimeter- and meter-wavelength facilities
either in commissioning or under construction that suggest optimism
for the future detection of extragalactic pulsars and radio
transients.  These instruments will provide higher raw sensitivities
(i.e., lower $S_{\mathrm{sys}}$) and/or larger bandwidths than we
have had available in this Arecibo search.  For specificity here, we
focus on single pulse searches, using Equation~(\ref{eq:trans}) as a
guide.
As discussed in \S\ref{s:periodicity}, detecting a periodic pulse
train at the distance of \objectname{M33} is likely to be possible
only for pulsars at the bright end of the pulsar pseudo-luminosity
function.  In contrast, with their much greater luminosities, giant
pulses should be detectable to much larger distances than
\objectname{M33} with higher sensitivity systems (or at the distance of
\objectname{M33} by less sensitive systems).

Table~\ref{tab:future} and Figure~\ref{fig:future} summarize the 
detection possibilities with future instruments.
Given the large number of facilities becoming available, we do not
attempt to summarize them all here.  Rather, we choose several
instruments as exemplars of future possibilities.  While our choice of
the specific instruments is somewhat arbitrary, there are multiple
instances in which various instruments have (or are expected to have)
similar performance.  For example, the system sensitivities of the
Expanded Very Large Array (EVLA) and the Green Bank Telescope (GBT)
are the same within~5\% ($S_{\mathrm{sys}} \approx 10$~Jy), and the 
Giant Metrewave Radio Telescope (GMRT; $S_{\mathrm{sys}} \approx 
11$~Jy) will offer a similar sensitivity once its new wide-band 
instrumentation (400 MHz) comes online. A future expansion of the 
Allen Telescope Array (ATA-256) has a system sensitivity that is within 
a factor of~2 of these telescopes.

In the near future, a substantial increase in sensitivity (by a factor $\sim$2--3)
that will be possible through the new instrumentation at Arecibo, such as the 
Mock spectrometers and the proposed Arecibo Ultimate Pulsar Processing
Instrument, offers significant promises for conducting renewed searches. 
Yet another promise is the proposed phased-array feed for Arecibo that will 
provide 42 beams instead of 7 ALFA beams; this will enable a much 
larger instantaneous sky coverage and therefore more efficient searches of 
extended objects such as M33. 

Furthermore, with an enormous increase in affordable computing power, 
it will become feasible to do searches with coherent dedispersion. This is
especially important when searching for giant pulses from Crab-like pulsars,
which are typically 1--10 \us\,in duration at 1.4 GHz, with the brighter ones being 
the narrower pulses \citep{bhatetal2008,ps07}. In such cases, the detection
sensitivity can be substantially boosted through matched filtering for very 
narrow pulses. This presents exciting, albeit computationally challenging, 
opportunities if there are Crab-like pulsars residing in external galaxies. 

As can be seen from Figure~\ref{fig:future}, the distance to which
giant pulses can be detected can be approximately normalized by
Arecibo.  Telescopes that have roughly 10\% of the sensitivity of
Arecibo (\hbox{EVLA}, MeerKAT) can probe to nearby Galactic
satellites, but probably cannot reach as far as \objectname{M33} (or
\objectname{M31}); telescopes comparable to Arecibo (\hbox{FAST}, SKA
Phase~1) can probe essentially the entire \objectname{Local Group};
and a telescope with $10\times$ the sensitivity of Arecibo (SKA
Phase~2) can reach well beyond the \objectname{Local Group}.

\begin{figure*}
\begin{center}
\includegraphics[angle=-90,width=0.95\textwidth]{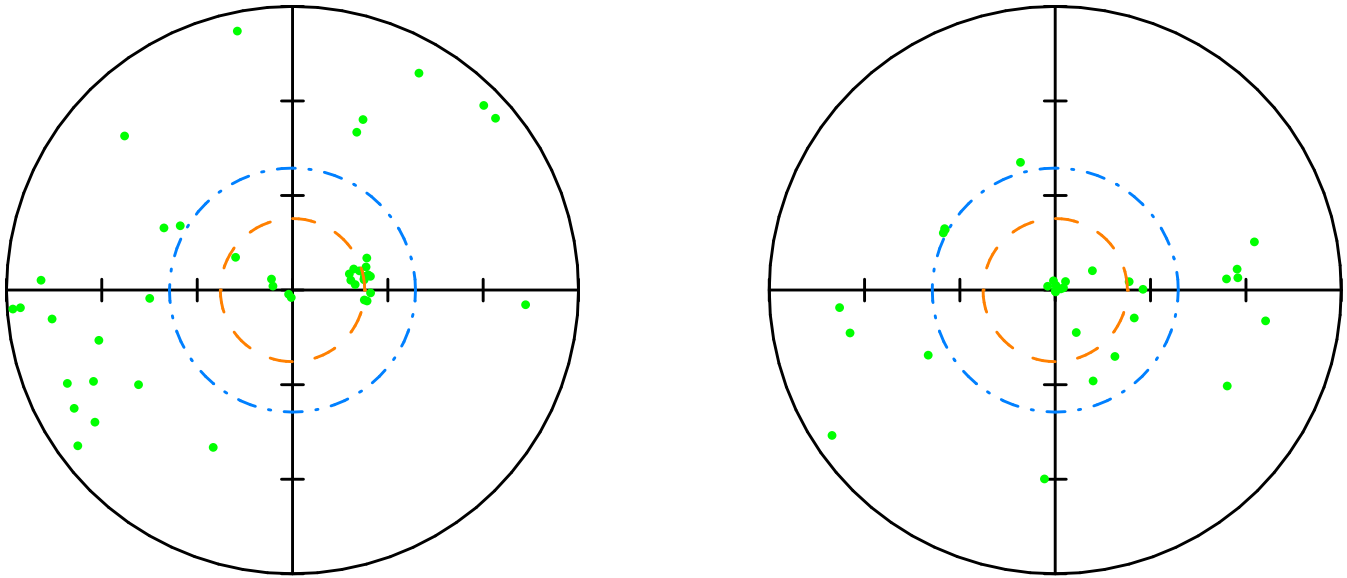}
\end{center}
\caption{Galaxies within 3 Mpc accessible to near-future radio telescopes for
searches for giant pulses, based on the nearby galaxy catalog of Karachentsev
et al. (2004), and assuming a Crab giant pulse of 20 kJy in peak flux density,
which will be detected with a S/N of 5 at M33Õs distance with AreciboÕs
sensitivity (Equation (2)). In both plots, the azimuth is right
ascension and galaxies are projected to the celestial equator along
lines of constant declination.  In both panels, the dashed orange circle shows the
distance to which a telescope having approximately 10\% of the
sensitivity of Arecibo (e.g., \hbox{EVLA}, \hbox{GBT}, ATA-256,
\hbox{MeerKAT}) could probe and the dot-dash blue circle shows the
distance to which an Arecibo-class telescope could probe.  The full
SKA could probe to an even greater distance than shown.
(\textit{Left}) Northern Hemisphere;
(\textit{Right}) Southern Hemisphere.}
\label{fig:LG}
\end{figure*}

We have used the nearby galaxy catalog of Karachentsev
et al. (2004) to assess
other potential targets; Figure~\ref{fig:LG} presents a visual
summary.  We define a potential target to be a galaxy at a distance
such that a giant pulse that is as bright as 20~kJy at the Crab's
distance ($\sim$2 kpc) will produce a signal-to-noise ratio of
at least 5 (Equation~\ref{eq:trans}).  In the northern (southern)
hemisphere, there are 21 (14) galaxies that are potential targets of
an Arecibo-class telescope; in both hemispheres, there are 10 galaxies
accessible to telescopes having approximately 10\% of the sensitivity
of Arecibo (e.g., \hbox{EVLA}, \hbox{GBT}, ATA-256, \hbox{MeerKAT}).
Of course, not all galaxies in Figure~\ref{fig:LG} will necessarily
host giant-pulse emitting pulsars.  Nonetheless, there are many more
targets to consider than just \objectname{M33}.

Despite the numerous challenges it poses, finding pulsars in other galaxies
will mark a major breakthrough, given the multi-pronged scientific benefits 
that will ensue. For instance, pulsars likely to be detected in such searches
will be young pulsars with high luminosities, hence their associations with 
any cataloged supernova remnants in the host galaxies may yield valuable 
information about the star formation rates as well as the pulsar birth rates 
and populations in those galaxies. Moreover, such extragalactic pulsars, 
through their dispersion, scattering and rotation measures, will also provide 
information about the  baryon density and magnetic field in the intervening 
intergalactic medium. With the sensitivity of the full SKA, it may even be 
possible build a partial census of pulsars in many nearby galaxies.

\section{Summary and Conclusions} \label{s:summary}

We have conducted an extensive and systematic search for pulsars and transient sources in M33 using Arecibo's seven-beam receiver system, ALFA. 
The galaxy, which is located at $\sim$0.95 Mpc and extends over roughly $73^{\prime} \times 45^{\prime}$ in the sky, was covered with 12 ALFA pointings, or 84 pixels in total, each of which was searched for 2--3 hr durations. 
We searched for both periodic and aperiodic signals, spanning up to 1000 \dmu\,in dispersion measure, and over timescales from $\sim$50 \us to several seconds.
While no signals of periodic or transient nature have emerged from these searches, they illustrate several underlying challenges and difficulties in carrying out such searches as well as improved efficiencies achievable through the availability of simultaneous multiple beams. 

For periodicity searches, our integration times of 2 hr translate to upper bounds of 3--10 \uJy in flux densities at M33's distance, which means very few pulsars  potentially detectable even in an optimistic scenario of M33 harboring a pulsar population akin to that in the Milky Way. 
For short-duration transients, the limiting peak flux densities from our analysis are $\Spk \la 20~{\rm mJy~W_{\rm ms}^{-1/2}}$ and the implied event rate limits correspond to an upper bound of about one Crab-pulsar-like object per ${\rm kpc^3}$ in M33. 
The limits on bright, one-off millisecond bursts \citep[e.g.,][]{lorimeretal2007} are not as constraining as those from the PALFA survey, however still significant and reinforce the conclusion of \citet{denevaetal2009}, i.e., the source population is unlikely to be homogeneous and isotropic in the sky.

Despite the numerous technical challenges inherent to such searches, the future looks promising. Several new radio facilities are currently under construction or in commissioning(e.g., LOFAR, FAST, MeerKAT); their higher raw sensitivities (i.e., lower \Ssys) and larger bandwidths mean more sensitive searches will be possible. 
Even the existing instruments such as EVLA, GBT and GMRT will offer opportunities in the near-term future with the availability of new instrumentation which will provide much larger bandwidths than currently possible (400--800 MHz). Arecibo itself will soon offer renewed opportunities, with a substantial increase in sensitivity (by a factor $\sim$2--3) that will be made possible through the new Mock spectrometers and other wideband instruments.  

In the longer term future, instruments such as the Five-hundred meter Aperture Spherical Telescope (FAST) and SKA Phase 1 will bring exciting opportunities for even more sensitive searches. While their sensitivities may be comparable to that of Arecibo, their sky coverages will be much larger and as a result they will essentially probe the entire Local group. 
Furthermore, with impressive technological advances being made in affordable supercomputing, searching with coherent dedispersion will eventually become feasible, which will substantially boost the detection sensitivity for giant pulses from Crab-like pulsars that may reside in the galaxies. Discoveries of extragalactic pulsars and transients will undoubtedly yield a multitude of scientific benefits, including understanding the populations of such objects in the host galaxies as well as serving as useful probes of the intervening intergalactic medium. 

\bigskip

\noindent
{\it Acknowledgments:}
We thank Arun Venkataraman for his assistance in transport of the large data volume to Swinburne University.  
This research was supported by NSF grant AST0807151 to Cornell University and an internal grant from Swinburne University. 
NDRB acknowledges the support received through the ANSTO travel grant 06/07-O-02 and a research grant from AAS, and thanks Matthew Bailes for continued support and encouragement extended to this project. MAM is supported by an Alfred P. Sloan Fellowship and by WVEPSCOR.
THH acknowledges partial support from NSF grant AST-0607492. 
We thank Willem van Straten for fruitful discussions and Sarah Burke-Spolaor for software assistance related to event analysis. 
Part of this research was carried out at the Jet Propulsion Laboratory, California Institute of Technology, under a contract with the National Aeronautics and Space Administration. We thank the referee, Scott Ransom, for providing comments that helped to improve
the clarity of some parts of the paper.
The Arecibo Observatory is operated by the National Astronomy and Ionosphere Center under a cooperative agreement to Cornell University.



\onecolumn

\begin{deluxetable}{llccccc}
\tablewidth{0pc}
\tablecaption{Survey pointings and observation parameters\label{obs}}
\tablehead{%
\colhead{Survey Field} & \colhead{MJD} & \colhead{R.A. (J2000)} & \colhead{DEC (J2000)} & 
\colhead{${\rm T _{obs}}$} & \colhead{$\Delta t _{samp}$ } & \colhead{Total time}  \\
 &  & &  & \colhead{(s)} & \colhead{($\mu$s)} & \colhead{(hr)} } 
\startdata
M33.1.A & 53985 &  01 33 49 & 30 59 52     &     3900   &        64  &        2.2 \\
M33.1.A & 53993 &  01 33 49 & 30 59 53     &     3900   &       100  & \\
M33.1.B & 53987 &  01 33 49 & 30 55 37     &     4000   &        64  &       3.3  \\
M33.1.B & 53995 &  01 33 49 & 30 55 36     &     4000   &       100  & \\
M33.1.B & 54074 &  01 33 50 & 30 55 36     &     4000   &       100  & \\
M33.1.C & 53989 &  01 33 49 & 30 51 20     &     4000   &        64  &        2.2\\
M33.1.C & 53997 &  01 33 49 & 30 51 21     &     4000   &       100  & \\
M33.2.A & 53985 &  01 34 17 & 30 44 21     &     3600   &        64  &        2.1\\
M33.2.A & 53993 &  01 34 17 & 30 44 21     &     3900   &       100  & \\
M33.2.B & 53987 &  01 34 17 & 30 40 05     &     4000   &        64  &       3.3\\
M33.2.B & 53995 &  01 34 17 & 30 40 05     &     4000   &       100  & \\
M33.2.B & 54075 &  01 34 18 & 30 40 05     &     4000   &       100  & \\
M33.2.C & 53989 &  01 34 18 & 30 35 49     &     3800   &        64  &        2.1\\
M33.2.C & 53997 &  01 34 17 & 30 35 49     &     3846   &       100  & \\
M33.3.A & 53986 &  01 33 23 & 30 47 20     &     3900   &        64  &        2.7\\
M33.3.A & 53994 &  01 33 23 & 30 47 21     &     4000   &       100  & \\
M33.3.A & 54073 &  01 33 22 & 30 47 21     &     1919   &       100  & \\
M33.3.B & 53988 &  01 33 22 & 30 43 06     &     4000   &        64  &        2.2\\
M33.3.B & 53996 &  01 33 22 & 30 43 06     &     4000   &       100  & \\
M33.3.C & 53992 &  01 33 23 & 30 38 50     &     3900   &        64  &        2.2\\
M33.3.C & 53998 &  01 33 23 & 30 38 49     &     4000   &       100  & \\
M33.4.A & 53986 &  01 33 49 & 30 29 51     &     3900   &        64  &       3.3 \\
M33.4.A & 53994 &  01 33 50 & 30 29 51     &     4000   &       100  & \\
M33.4.A & 54074 &  01 33 49 & 30 29 52     &     3800   &       100  & \\
M33.4.B & 53988 &  01 33 50 & 30 25 35     &     3544   &        64  &        2.1\\
M33.4.B & 53996 &  01 33 50 & 30 25 36     &     4000   &       100  & \\
M33.4.C & 53992 &  01 33 49 & 30 21 19     &     3800   &        64  &        2.2\\
M33.4.C & 53998 &  01 33 49 & 30 21 20     &     4000   &       100  & \\
\enddata
\tablecomments{The coordinates given for each pointing are for the central ALFA beam.}
\end{deluxetable}

\begin{deluxetable}{lccc}
\tablewidth{0pc}
\tablecaption{Statistics of RFI and candidate events\label{stats}}
\tablehead{
\colhead{Quantity} & \multicolumn{2}{c}{Matched filtering} & \colhead{Time domain clustering} \\
\colhead{} & \colhead{Short events} & \colhead{Long events} & \colhead{} }
\startdata
 & \multicolumn{3}{c}{RFI events (classification 4)} \\
\cline{2-4}
No. of data segments &  143      & 40 &  153   \\
No. of events     &   207  & 83 &      375     \\
\% of observing time &  0.06 & 0.024 &  0.11 \\
\hline
 & \multicolumn{3}{c}{Interesting events (classification 1 or 2)} \\
\cline{2-4}
No. of data segments &      72 & 16 &         23    \\
No. of events &     124  &18 &       27     \\
\% of observing time & 0.035 & 0.005 & 0.008 \\
\enddata
\tablecomments{The classification into ``short" and ``long" is based on the event timescale being \la 10 ms and \ga 10 ms, respectively.
``Data segment" corresponds to a 4-s chunk around the event time; often multiple events are detected within a single 4-s chunk, 
hence the total number of events larger than the number of data segments.}
\end{deluxetable}

\begin{deluxetable}{lcccc}
\tablewidth{0pc}
\tablecaption{Future Searches for Extragalactic
	Pulses\label{tab:future}}
\tablehead{%
\colhead{Telescope} & \colhead{$\nu$} & \colhead{$S_{\mathrm{sys}}$} 
	& \colhead{$\Delta\nu$} & \colhead{$(S/N) _{\rm GP}$} \\
                    &                 & \colhead{(Jy)}
	& \colhead{(MHz)}       & \colhead{(at ${\rm D_{M33}}$ )}}
\startdata
Arecibo & 1.4~GHz &  3.0 & 400 & 10 \\
LOFAR   & 0.2~GHz & 19.1     & 32  &  8 \\
EVLA    & 1.4~GHz & 10.4 & 500 & 3.2 \\
GMRT   & 1.2~GHz & 11.0 & 400 & 2.7 \\
FAST    & 1.4~GHz &  2.2 & 400 & 13.6 \\
MeerKAT & 1.4~GHz & 17.8 & 500 & 1.9\\
SKA Phase~1 & 1.4~GHz & 3.6 & 500 & 9.3 \\
SKA Phase~2 & 1.4~GHz & 0.3 & 1000 & 160 \\
\enddata
\tablecomments{The instruments listed are the Arecibo Observatory, the Low Frequency
Array (LOFAR), the Expanded Very Large Array (EVLA), the Giant Metrewave
Radio Telescope (GMRT), the Five-hundred-meter Aperture Spherical
Telescope (FAST), the Karoo Array Telescope (MeerKAT), the SKA Phase 1
(SKA-1), and the SKA Phase 2 (SKA-2). As many of the telescopes are still under
construction or have just started commissioning, parameters listed should be
considered indicative but not definitive. 
$(S/N) _{\rm GP}$ denotes the predicted signal-to-noise ratio (at M33Õs distance, 
${\rm D_{M33}}$) for a giant pulse that is as bright as
$\sim$20 kJy at the CrabÕs distance ($\sim$2 kpc).}
\end{deluxetable}

\end{document}